\documentclass{article}

\PassOptionsToPackage{numbers}{natbib}

\usepackage[final]{nips_2016}

\usepackage[utf8]{inputenc} 
\usepackage[T1]{fontenc}    
\usepackage{hyperref}       
\usepackage{url}            
\usepackage{booktabs}       
\usepackage{amsfonts}       
\usepackage{amsmath}
\usepackage{nicefrac}       
\usepackage{microtype}      
\usepackage{color}

\DeclareMathOperator*{\argmin}{arg\,min}

\title{Partially blind domain adaptation for age prediction from DNA methylation data}

\author{
	Lisa Handl \\
	Max Planck Institute for Informatics \\
	Saarland Informatics Campus \\
	66123 Saarbrücken \\
	\texttt{lisa.handl@mpi-inf.mpg.de}
	\And
	Adrin Jalali \\
	Max Planck Institute for Informatics \\
	Saarland Informatics Campus \\
	66123 Saarbrücken \\
	\texttt{ajalali@mpi-inf.mpg.de}
	\And
	Michael Scherer \\
	Max Planck Institute for Informatics \\
	Saarland Informatics Campus \\
	66123 Saarbrücken \\
	\texttt{mscherer@mpi-inf.mpg.de}
	\And
	Nico Pfeifer \\
	Max Planck Institute for Informatics \\
	Saarland Informatics Campus \\
	66123 Saarbrücken \\
	\texttt{nico.pfeifer@mpi-inf.mpg.de}
}

\begin{document}
	
	\maketitle
	
	\begin{abstract}
		Over the last years, huge resources of biological and medical data have become available for research. This data offers great chances for machine learning applications in health care, e.g. for precision medicine, but is also challenging to analyze. Typical challenges include a large number of possibly correlated features and heterogeneity in the data. One flourishing field of biological research in which this is relevant is epigenetics. Here, especially large amounts of DNA methylation data have emerged. This epigenetic mark has been used to predict a donor’s ``epigenetic age'' and increased epigenetic aging has been linked to lifestyle and disease history. In this paper we propose an adaptive model which performs feature selection for each test sample individually based on the distribution of the input data. The method can be seen as partially blind domain adaptation. We apply the model to the problem of age prediction based on DNA methylation data from a variety of tissues, and compare it to a standard model, which does not take heterogeneity into account. The standard approach has particularly bad performance on one tissue type on which we show substantial improvement with our new adaptive approach even though no samples of that tissue were part of the training data.
	\end{abstract}
	
	\section{Introduction}
	\label{sec:intro}
	
	Epigenetics, the heritable modification of phenotypes that is not encoded by DNA, has become an important field in biological research. The best-studied epigenetic mark is DNA methylation, which was detected to play a role in long-term repression of genes through promoter methylation, X-chromosomal inactivation and genomic imprinting \cite{reviewSchub}. It refers to the covalent addition of  methyl groups to the C5 position of cytosines, predominately found in CpG dinucleotides. Due to the growing number of datasets in this field, a connection between the methylation pattern of genomic DNA and its donor's chronological age was reported \cite{heynAge,agingBell,driftTeschendorff}. On this basis, several studies created models to predict chronological age from DNA methylation data \cite{horvath2013dna,hannum,florath}. They defined the outcome of the prediction as the ``epigenetic age'' of the person and linked increased epigenetic aging to lifestyle factors and disease history. As a concept of biological age, the epigenetic age is more informative about the individual's health status than chronological age and can be useful to optimize disease treatment.
	
	Due to the large number of sometimes strongly correlated features, DNA methylation data at the CpG level is challenging to model. Ordinary least squares regression leads to predictors with large variance because a large positive coefficient of one variable can be compensated by a large negative coefficient of a correlated variable. One way to prevent this is to use feature selection, e.g., by penalizing the $L_1$ norm of the coefficient vector in the loss function (LASSO). This type of regularization will set many coefficients to zero, leading to sparse and more robust models. An alternative approach is ridge regression, which penalizes the $L_2$ norm instead. Ridge regression forces coefficients to be small, but does not strictly set them to zero. In the presence of correlated features, ridge regression averages the coefficients while LASSO tends to pick one of the correlated variables. The elastic net penalizes a linear combination of the $L_1$ and $L_2$ norm of the coefficients and has been proposed to combine the advantages of LASSO and ridge regression~\cite{zou2005regularization}. It still performs feature selection, but tends to average the coefficients of included correlated features in a similar way as ridge regression.
	
	Another difficulty, which is present in many biological and medical datasets, is the heterogeneity of the data. Small differences in data acquisition and processing (e.g., different protocols in laboratories or standards in clinics) may lead to biases and make it hard to compare data from different sources. Domain adaptation attempts to correct for mismatches between distributions in scenarios where large amounts of data from a source domain and small amounts of data from a target domain are available~\cite{schweikert2009empirical}. An even harder problem is blind domain adaptation, where data from the target domain is not available at training time~\cite{uzair2016blind}.
	
	In this paper, we present an approach which performs feature selection for each test sample individually to reduce effects of data heterogeneity. We build on ideas from~\cite{jalali2016} to find features that behave similarly in training and test data, but do not use a predefined set of weak learners. Instead, we train a full model for each test sample. Since the models are still trained only on the training data, but information from the test samples is used to select appropriate features, our setting can be seen as partially blind domain adaptation. We apply the method to the problem of age prediction based on a large DNA methylation dataset. The main source of heterogeneity in this data comes from the use of different tissues, some of which are not present in our training data. We show that our approach leads to improved test errors for samples from the cerebellum of the human brain, which is the tissue in our data that leads to the largest errors with standard models that do not account for the bias.

	\section{Methods}
	\label{sec:methods}
	
	The core idea of our approach is to train test sample-specific models, considering only features in which we have high confidence for the test sample at hand. In a large heterogeneous dataset, it is possible that only some features cause the heterogeneity while others behave similarly in training and test data. Obviously, features that behave very differently should not be used in a predictive model. Excluding them and relying only on similarly behaving features can thus lead to a more robust model. 
	
	This can be expressed more formally in the framework of domain adaptation. Assume that the training and test samples are drawn independently from two joint probability distributions ${P_S(X,Y)} = {P_S(Y \mid X})\cdot {P_S(X)}$ and $P_T(X,Y) = P_T(Y \mid X) \cdot P_T(X)$, respectively. Here $S$ stands for source domain and $T$ for target domain. A classical assumption in domain adaptation is that the conditional distributions, $P_S(Y \mid X) = P_T(Y \mid X)$, are the same in source and target domain while the distributions of input features may be different, i.e., $P_S(X) \neq P_T(X)$. This setting is called the covariate shift case. We weaken the covariate shift assumption by requiring equal conditional distributions only for part of the available features. More precisely, we assume that there is a subset $M \subset \{1, \ldots, m\}$ of all features on which the same model can accurately predict the outcome from training and test inputs. This means that $P_S(Y \mid X_M) = P_T(Y \mid X_M)$, where $X_M$ denotes the subvector of the random vector $X$ containing only features in the reduced feature set $M$. The distribution of input features as well as the relationship between $Y$ and the remaining features may be different in source and target domain, i.e., $P_S(X) \neq P_T(X)$ and $P_S(Y \mid X_N) \neq P_T(Y \mid X_N)$ for $N=\{1, \ldots, m\} \setminus M$. In addition, we allow that $M$, the set of features that behave similarly in predicting $Y$, may be different for different test samples. Thus, a good choice of $M$ has to be determined for each test sample separately.	
	
	For this purpose, we propose a model-based approach to estimate a confidence of each feature for a given test sample. We then train a full model for each test sample, learning from the training data and using only high-confidence features. Since we do not know the response variable $Y$ for the test samples, we explore the dependency structure within $X$ to determine confidences. The underlying assumption is that if there is a subset of features, $M$, whose dependency structure is very similar in training and test data, then the relationship between $Y$ and these features will also be similar in training and test data. More formally, writing $X_f$ for the value of feature $f$ and $X_{-f}$ for the values of all other features, we assume that if $P_{S}(X_{f}|X_{-f}) \approx P_{T}(X_{f}|X_{-f})$ holds for all features $f\in M$, then $P_{S}(Y|X_{M}) \approx P_{T}(Y|X_{M})$.

	\paragraph{Model types}
	
	We apply two main model types in this paper: elastic net and Gaussian process models. The elastic net is a form of regularized linear regression, which penalizes a combination of the $L_1$ and $L_2$ norm of the coefficient vector~\cite{zou2005regularization}. More precisely, it finds
	\begin{equation*}
	\hat\beta = \argmin_{\beta} \left( \frac{1}{2n}\|y - \boldsymbol{X}\beta\|_2^2 + \lambda\left( \alpha \|\beta\|_1 + \frac{1-\alpha}{2} \|\beta\|_2^2 \right)\right),
	\end{equation*}
	where $\boldsymbol{X},y$ is the training data and $n$ is the number of samples that it contains. While $\alpha \in [0,1]$ determines the mixing ratio of $L_1$ and $L_2$ penalty and is often set to a fixed value, $\lambda \geq 0$ controls the strength of regularization and is usually determined using cross-validation. Gaussian process models are a type of non-parametric Bayesian regression, where the prior distribution over regression functions is a Gaussian process with mean zero and a covariance function which is typically specified in the form of a kernel~\cite{rasmussen2006gaussian}. Bayesian models have the advantage that they provide not only a predicted value, but a distribution of possible output values for any new input. In the setting applied in this paper this distribution is Gaussian and known explicitly.

	\paragraph{Datasets}
	
	We collected 26 datasets from the Gene Expression Omnibus (GEO, \href{https://www.ncbi.nlm.nih.gov/geo/}{ncbi.nlm.nih.gov/geo}) and the Cancer Genome Atlas (TCGA, \href{http://cancergenome.nih.gov/}{cancergenome.nih.gov}), which analyzed DNA methylation by the Illumina Infinium HumanMethylation450 BeadChip. Then, we combined these datasets using RnBeads \cite{RnBeads2014} and split it into a training and test set consisting of 1866 and 1007 samples, respectively. All samples included were obtained only from healthy tissues. The training set contains 16 and the test set 6 different tissues, with a focus on blood samples for both sets. For the training set, samples from donors with chronological ages between 0 and 103 years were used. The age range for the test set is 0-70 years, accordingly. SNP-removal, removal of gonosomal CpGs and data normalization with the BMIQ method \cite{BMIQ} were performed by RnBeads. We reduced the initial number of features from 466,094 to 12,980 features using an elastic net model with strong regularization ($\lambda = 1.1\cdot 10^{-4}$). This is necessary for computational reasons since we train a very large number of models.

	\paragraph{Reference model}
	
	We used a similar type of model as baseline as presented in~\cite{horvath2013dna}, namely, an elastic net model with $\alpha=0.8$, followed by least squares linear regression based on the selected features. This model has been trained on our training dataset and the regularization parameter $\lambda$ has been selected via 10-fold cross-validation.
	
	\paragraph{Adaptive model}
	
	To estimate confidences of the features of test samples, we first trained a Gaussian process model for each feature, based on all other features. We chose a linear kernel and additive Gaussian noise, and determined the kernel parameter and noise variance of each model using marginal likelihood maximization. 
	For a given test sample, $X_i$, these models can be used to predict a posterior distribution of $X_{i,f}$ (the value of $X_i$ for some feature $f$), given the values of all other features, which we denote by $X_{i,-f}$. In our setting, we obtain a Gaussian posterior distribution, $\mathrm{N}(\mu_{g_f}(X_{i,-f}), \sigma^2_{g_f}(X_{i,-f}))$. By comparing the observed value, $X_{i,f}$, to the predicted distribution, we can quantify how well $X_{i,f}$ fits to what is expected according to the training data. We quantify the confidence of feature $f$ for $X_i$ as proposed in~\cite{jalali2016} by	
	\begin{equation}
	c_f(X_i) = 2 \cdot \Phi\left(-\left|\frac{X_{i,f} - \mu_{g_f}(X_{i,-f})}{\sigma_{g_f}(X_{i,-f})}\right|\right),
	\end{equation}
	where $\Phi$ denotes the cumulative distribution function of the standard normal distribution. This can be interpreted as the probability that a value like $X_{i,f}$ or more extreme occurs according to its predicted distribution. After estimating confidences for all test samples and features, we use this information to train an age predictor for each test sample individually, based on only its high-confidence features. Here we used the same model type as for the reference model described in the previous paragraph, but only 3-fold cross-validation. We tried multiple thresholds for defining high-confidence features, choosing the top 10\%, 20\%, 30\% or 40\% for each test sample. Note that the confidence estimation (and feature selection) is specific to the test sample, but each model is trained on the same training data. Moreover, no information on the output of test samples is used. 
	
	The adaptive model is computationally expensive since it involves fitting a large number of models. If $m$ is the number of features and $k$ is the number of test samples, then $m+k$ models are fitted in total. 
	However, each of the main steps (i.e., fitting $m$ models for confidence estimation and fitting $k$ final models) can easily be parallelized to speed up computations.

	\section{Results and discussion}
	\label{sec:results}
	
	\paragraph{Reference model} 
	
	We trained the reference model on the training dataset with 12,980 features. The optimal regularization parameter determined by cross-validation is $\lambda = 0.01$, which corresponds to 436 features with nonzero coefficients. Table~\ref{tab:reference} shows the mean and median absolute test errors for the full test dataset and for cerebellum samples separately. We obtained a mean absolute error of 4.82 on the full test dataset. Given the wide range of ages and tissues considered, an error of this size seems reasonable. For cerebellum samples, however, we obtained a mean absolute error of 16.95, which is more than three times larger. This is not surprising as cerebellum samples are not present in our training data, but much larger than desirable. Both for the full test dataset and for cerebellum samples, the median absolute error is slightly lower than the mean.
	
	\begin{table}
		\caption{Mean and median absolute test errors of the reference model for the full test dataset and for cerebellum (CRBM) samples.}
		\label{tab:reference}
		\centering
		\begin{tabular}{llrrrr}
			\toprule
			\multicolumn{2}{c}{Type of test error} & Test error \\
			\midrule
			Full test dataset 	& mean 		& 4.82	\\
			& median  	& 3.45	\\
			CRBM samples		& mean		& 16.95	\\
			& median	& 16.57	\\
			\bottomrule
		\end{tabular}
	\end{table}

	\paragraph{Adaptive model}
	
	\begin{table}
		\caption{Mean and median absolute test errors of the adaptive model for the full test dataset and for cerebellum (CRBM) samples.}
		\label{tab:adaptive}
		\centering
		\begin{tabular}{llrrrr}
			\toprule
			& & \multicolumn{4}{c}{Percentage of high-confidence features} \\
			\cmidrule{3-6}
			\multicolumn{2}{c}{Type of test error}	
			& Top 10\%	& Top 20\%	& Top 30\%	& Top 40\%	\\
			\midrule
			Full test dataset 	& mean 		& 7.96     	& 6.61		& 6.16		& 5.78 \\
			& median  	& 6.82		& 5.69		& 4.87		& 4.30 \\
			CRBM samples		& mean		& 12.78		& 12.96		& 13.36		& 14.11	\\
			& median	& 10.19		& 12.63		& 13.78		& 14.94 \\
			\bottomrule
		\end{tabular}
	\end{table}
	
	In addition, we trained the adaptive model described in Section~\ref{sec:methods} for different thresholds defining high-confidence features. The resulting mean and median absolute test errors are presented in Table~\ref{tab:adaptive}. For cerebellum samples, each of the adaptive models gave lower errors than the reference model. The performance on cerebellum samples is best when only features with the top 10\% of confidences are used, leading to a mean absolute error of 12.78 and an even lower median of 10.19. When increasing the threshold, the errors on cerebellum samples slowly become larger, but still stay well below the corresponding errors of the reference model. These results demonstrate that restricting the model to high-confidence features can reduce the error on samples for which a distribution mismatch with the training data is present. A stronger restriction, which corresponds to a stronger focus on high confidences, leads to a larger improvement. At the same time, the errors on the full test dataset are larger for the adaptive models than for the reference model. Here we observe the opposite development. Errors decrease continuously with increasing threshold, from 7.96 for a threshold of 10\% to 5.78 for a threshold of 40\% in the case of mean absolute error. This can be explained by the fact that if all features behave the same way for training and test data, selecting only the ``best'' of them will not lead to an improvement. Thus, if no distribution mismatch is present, restricting the model to far less features than the reference model is expected to lead to increased errors. Despite this, all errors on the full test dataset are still below the errors on cerebellum samples.

	\section{Conclusions and outlook}
	\label{sec:conlcusions}
	
	Heterogeneous data is ubiquitous in applications of machine learning in biology and medicine. In this paper we analyzed a large dataset of DNA methylation, which is heterogeneous because it was derived from multiple tissues. We proposed an adaptive model for predicting the donor's chronological age from this data. For each test sample the model selects features according to which the test sample behaves in a similar way as the training data. Then, it uses only these reliable features for prediction. Our model performs better than a non-adaptive reference model on samples from the cerebellum of the human brain. This tissue was not represented in the training data and lead to the largest errors in the reference model. Thus, we demonstrated that our approach to partially blind domain adaptation can be a powerful way to reduce test errors on samples that are different from the training data. This improvement has a price when applying the model to test samples with the same or a very similar distribution as the training data. The main reason is that strictly excluding features restricts the model, which is not beneficial if no distribution mismatch is present. Of course, these findings need to be verified on additional datasets. 
	
	One possibility for improvement of the proposed model might be to weight features according to their confidences instead of including or excluding them strictly. This might improve the performance on samples without a distribution mismatch and will be subject of future work.

	\subsubsection*{Acknowledgments}
	
	This work was prepared within the project \emph{XplOit} of the initiative "i:DSem – Integrative Datensemantik in der Systemmedizin", which is funded by the German Federal Ministry of Education and Research (BMBF).
	
	\small
	
	\bibliography{biblio}
	
\end{document}